\pdfoutput=1
\documentclass{PoS}
\def\rQCED{{\rm QCED}}
\newcommand{\KK}{${\cal KK}$}
\title{Recent IR-Improved results for LHC/FCC physics}

\ShortTitle{Recent IR-Improved results for LHC/FCC physics}

\author{\speaker{B.F.L. Ward}\\%
        Baylor University\\
        E-mail: \email{bfl\_ward@baylor.edu}}

\author{Aditi Mukopadhyay\\
           AstraZeneca \\
       E-mail: \email{aditi.banya@gmail.com}}
       
  \author{Bahram Shakerin \\
           Baylor University \\
       E-mail: \email{bahram\_shakerin@baylor.edu}}     
       
       \author{ Zbigniew A. W\c{a}s\\
  Institute of Nuclear Physics Polish Academy of Sciences\\
       E-mail: \email{z.was@cern.ch}}
       
       \author{Scott A. Yost \\
        The Citadel \\
       E-mail: \email{yosts1@citadel.edu}}

\abstract{ We present recent developments in the theory and application of IR-improved QED$\otimes$QCD resummation methods, realized by MC event generator methods, 
for LHC and FCC physics scenarios.}
\centerline {\bf BU-HEPP-17-03, Dec., 2017}
\FullConference{13th International Symposium on Radiative Corrections (Applications of Quantum Field Theory to Phenomenology)\\
		 25-29 September, 2017 \\
		 St. Gilgen, Austria}

\begin{document}

\section{Introduction}
    The original discussion between one of us (BFLW) and Prof. F. Berends in the 1988 ICHEP-Munich Conference Dinner on the accuracy one could expect from two methods of resummation in the SM EW theory that were being pursued at that time, the Jackson-Scharre~\cite{js} approach and the YFS~\cite{yfs} approach, for a given level of exactness, has today, in the context of QCD, been extended to include another component: the issue of a hard cut-off for the IR versus resummed IR integrability. Ultimately, the precision data should be able to settle this extended component as it resolves the original discussion for the QCD case. In this context, we note that, following the discovery of the BEH~\cite{beh}  boson by ATLAS~\cite{atlas-hggs} and CMS~\cite{cms-higgs}, we have entered the era of QCD with precision tags $\lesssim 1.0\%$ with accompanying EW precision tags at the per mille level for processes such as single heavy gauge boson production at the LHC. \par
    In addressing the present paradigm in precision physics at the LHC, and its implications for the futuristic FCC, we have pursued exact amplitude-based resummation realized on an event-by-event basis via shower/matrix element(ME) matched MC's. In this way, we have achieved enhanced precision for a given level of exactness: LO, NLO, NNLO, $\ldots$ . Currently, in the Herwig6.5~\cite{hwg} environment we have a realization of IR-improved parton showers in the MC Herwiri1.031~\cite{hwri} by two of us 
(BFLW and SAY) which is elevated to the exact NLO shower/ME matched level via the MC@NLO~\cite{mcnlo} and via the MG5\_aMC@NLO~\cite{mg5aMC} frameworks as MC@NLO/Herwiri1.031~\cite{mcnlo-hwri} and MG5\_aMC@NLO/Herwiri1.031~\cite{mg5_amc-hwri}, respectively, by three of us (AM, BFLW and SAY). In the Pythia8~\cite{pythia8} environment, we have the realization of IR-improved (IRI) Pythia8~\cite{cpc-py8} by one of us (BFLW)
with its corresponding NLO shower/ME matched  MG5\_aMC@NLO/IRI-Pythia8. More recently, in the new MC {\KK}MC-hh~\cite{kkmchh} some of us (BFLW, ZAW and SAY) we have realized exact ${\cal O}(\alpha^2L)$ CEEX EW corrections in a hadronic MC in the Herwig6.5 environment. \par
    From Refs.~\cite{hwri,mcnlo-hwri} some of us (AM,BFLW,SAY) have shown that IR-improvement in Herwig6.5 via Herwiri1.031 leads to improved precision in both the central $|\eta_\ell|\lesssim 2.5$ region for the ATLAS, CMS, D0 and CDF data and in the more forward region of LHCb where $2.0<\eta_\ell<4.5$. Here, $|\eta_\ell|$ is the lepton pseudorapidity in respective single $Z/\gamma^*$ production production with decay to lepton pairs. One of us (BFLW) has shown~\cite{iri-semi-an} that the IR-improved semi-analytical paradigm is available for the processes just mentioned. In what follows, we extend our methods to the analysis of LHC W+ n jets data, to the FCC discovery physics and to the interplay of IR-improved parton showers with exact ${\cal O}(\alpha^2L)$ CEEX EW corrections in {\KK}MC-hh.\par
    We are moved to note that this year is the 50th anniversary of the seminal paper by S. Weinberg~\cite{weinbg1} in which he formulated his foundational model of leptons in creating the spontaneously broken $SU_{2L}\times U_1$ EW theory~\cite{slg,aslm}, one of the key components of the SM. The progress on precision theory has been essential to the establishment of the SM\cite{weinbg1,slg,aslm,gw,pltzr}. As we celebrate 50 years of the SM~\cite{50yr-SM}, we are also obliged to look to the future with the FCC~\cite{fcc-study} on the horizion, 
\begin{figure}[t]
\begin{center}
\includegraphics[width=3.5in]{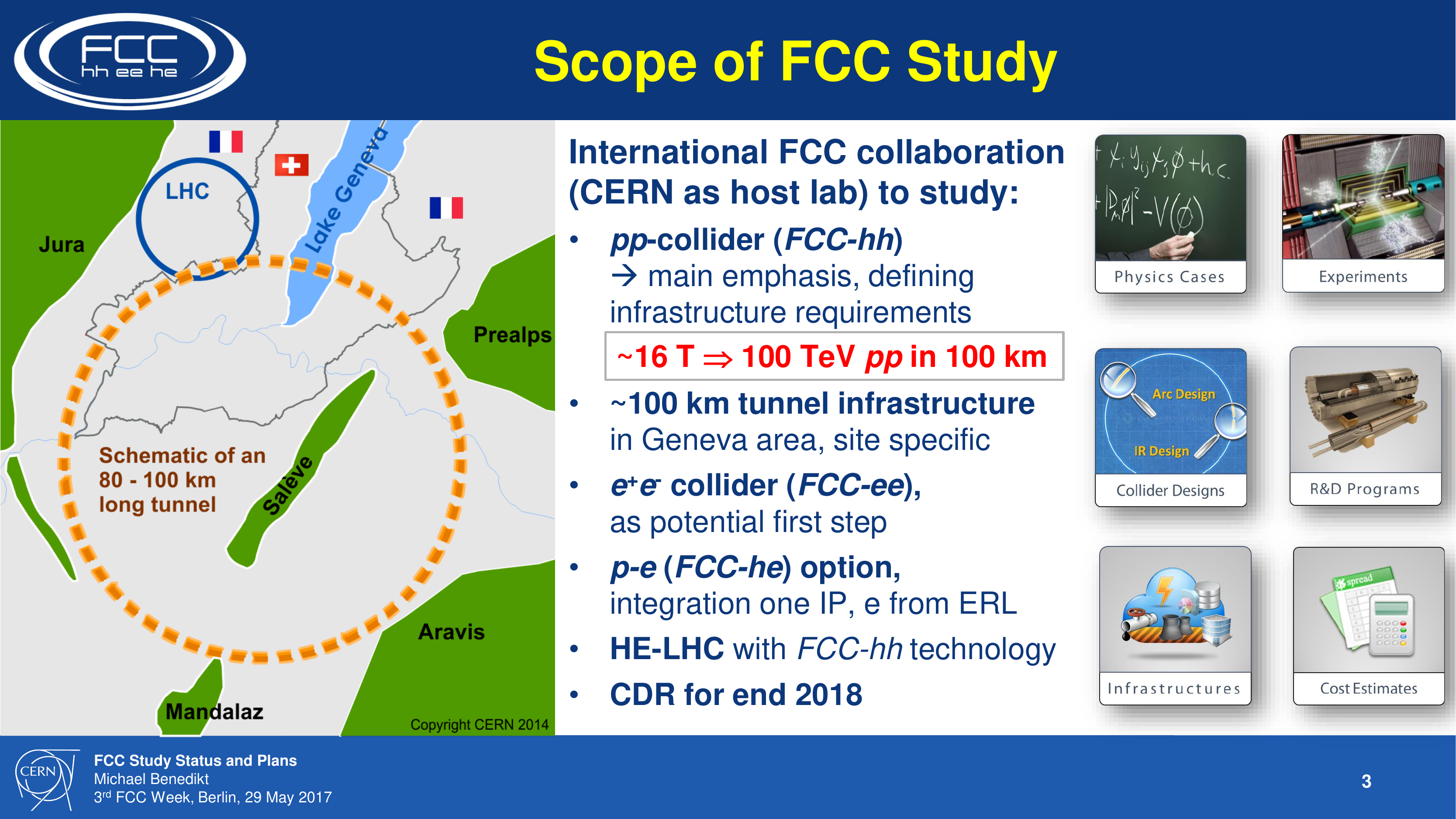}
\end{center}
\caption{\baselineskip=7pt FCC device planned for the future at CERN as depicted in Ref.~\cite{fcc-study}.}
\end{figure}
which will feature a 100 TeV hadron collider and tera-Z $e^+e^-$ colliding beam device. The future success of the latter devices will depend strongly on the  progress of precision theory as we discuss in this meeting.\par
    The paper is organized as follows. In the next section we briefly review the parton shower implementation of exact amplitude-based resummation theory.  In section 3, we turn to the interplay of IR-improved DGLAP-CS QCD theory and shower/ME matched precision via comparisons with LHC data on $W + n$ jets and via predictions for FCC discovery physics. In Section 4, we discuss the interplay of IR-improved DGLAP-CS QCD theory and exact ${\cal O}(\alpha^2 L)$ CEEX EW corrections in single $Z/\gamma^*$ production at the LHC. We sum up in Section 5.
    
\section{Review of Parton Shower Implementation of Exact Amplitude-Based Resummation Theory}
In this section we briefly recapitulate the elements of the parton shower implementation of exact amplitude-based resummation theory. The starting point is the master formula
\begin{eqnarray}
&d\bar\sigma_{\rm res} = e^{\rm SUM_{IR}(QCED)}
   \sum_{{n,m}=0}^\infty\frac{1}{n!m!}\int\prod_{j_1=1}^n\frac{d^3k_{j_1}}{k_{j_1}} \cr
&\prod_{j_2=1}^m\frac{d^3{k'}_{j_2}}{{k'}_{j_2}}
\int\frac{d^4y}{(2\pi)^4}e^{iy\cdot(p_1+q_1-p_2-q_2-\sum k_{j_1}-\sum {k'}_{j_2})+
D_\rQCED} \cr
&{\tilde{\bar\beta}_{n,m}(k_1,\ldots,k_n;k'_1,\ldots,k'_m)}\frac{d^3p_2}{p_2^{\,0}}\frac{d^3q_2}{q_2^{\,0}},
\label{subp15b}
\end{eqnarray}
\small
where {\em new} (YFS-style) {\em non-Abelian} residuals 
{$\tilde{\bar\beta}_{n,m}(k_1,\ldots,k_n;k'_1,\ldots,k'_m)$} have {$n$} hard gluons and {$m$} hard photons. The infrared functions in (\ref{subp15b}) are defined as follows:
\begin{eqnarray}
{\rm SUM_{IR}(QCED)} &=&2\alpha_s\Re { B^{nls}_{QCED}}+2\alpha_s{{\tilde B}^{nls}_{QCED}}\cr
{ D_\rQCED} &=&\int \frac{d^3k}{k^0}\left(e^{-iky}-\theta({K_{max}}-k^0)\right){{\tilde S}^{nls}_{QCED}}
\label{irfns}
\end{eqnarray}
where {$K_{max}$} is a``dummy'' in that (\ref{subp15b}) does not depend upon it and{\baselineskip=12pt
\begin{eqnarray}
{ B^{nls}_{QCED}} &\equiv& { B^{nls}_{QCD}{+{\footnotesize\frac{\alpha}{\alpha_s}}}B^{nls}_{QED}},\cr
{{\tilde B}^{nls}_{QCED}}&\equiv& {{\tilde B}^{nls}_{QCD}{+{\footnotesize\frac{\alpha}{\alpha_s}}}{\tilde B}^{nls}_{QED}}, \cr
{{\tilde S}^{nls}_{QCED}}&\equiv& {{\tilde S}^{nls}_{QCD}{+}{\tilde S}^{nls}_{QED}}.
\end{eqnarray} 
Here, { ``nls''} denotes { DGLAP-CS synthesization} as explained in Ref.~\cite{dglpsyn}.
\par
Shower/ME matching results in the replacements {$\tilde{\bar\beta}_{n,m}\rightarrow \hat{\tilde{\bar\beta}}_{n,m}$}. In this context, starting from the basic formula
\begin{equation}
{d\sigma} =\sum_{i,j}\int dx_1dx_2{F_i(x_1)F_j(x_2)} d\hat\sigma_{\rm res}(x_1x_2s),
\label{bscfrla}
\end{equation}
the connection to MC@NLO proceeds as explained in Ref.~\cite{mcnlo-hwri}. A similar connection to POWHEG~\cite{pwhg} is possible~\cite{bflwtoapp}.\par
What one can see from the connection to MC@NLO given in Ref.~\cite{mcnlo-hwri} is that the relationship between the hard gluon (and photon) residuals and the exact NLO (NNLO) corrections implies that the study of the theoretical precision of $d\sigma$ in (\ref{bscfrla}) necessarily entails the study of the precision of the exact NLO (NNLO) corrections.
This immediately raises the question of the interpretation of the divergences in the latter corrections as they are regulated to +-functions. Resummation renders these divergences 
integrable. To see how this integrability is manifested in observations, we turn first to Drell-Yan processes with LHC data on $W+ n$ jets and before turning to FCC single $Z/\gamma^*$ discovery physics to probe another process and phase space regime.\par
\section{Interplay of IR-Improved DGLAP-CS Theory and Exact NLO ME Matched Shower Precision: Comparison with LHC $W + n$ Jets Data }

We make comparisons in this Section between the LHC data on $W+ n$ jets, n=1,2,3, and the exact NLO ME matched QCD parton shower predictions in the MG5\_aMC@NLO framework with the parton shower realized via Herwig6.5 and Herwiri1.031 respectively for the unimproved and IR-improved results. We focus on the $p_T$ spectra here and refer the reader to Refs.~\cite{bsh1,bsh2} for more complete studies  involving other observables.\par
We start with the analysis~\cite{bsh1} by two of us (BS and BFLW) of the ATLAS~\cite{atlas-7-TeV} W+ 1 jet data at 7 TeV. We show in Fig.~\ref{figp2} the comparison between the data for the leading jet $p_T$ and the predictions by  MG5\_aMC@NLO/Herwig6.5, labeled as 'herwig' in the figure, and by  MG5\_aMC@NLO/Herwiri1.031, labeled by 'herwiri' in the figure.
The Herwig6.5 simulations include an intrinsic  Gaussian transverse momentum distribution with a root-mean-square value of 2.2 GeV/c.
\begin{figure}[h]
\begin{center}
\includegraphics[width=3.5in]{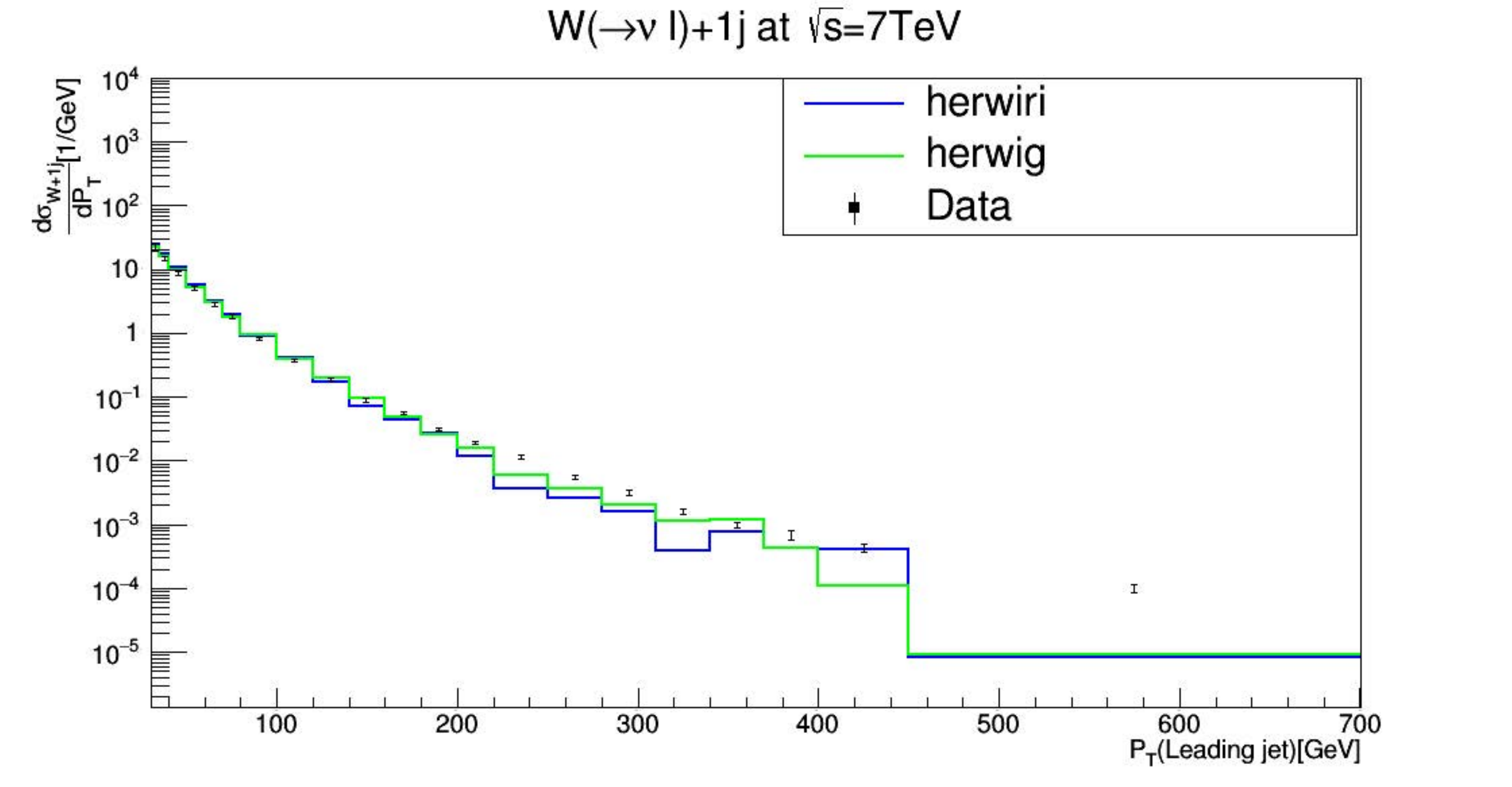}
\end{center}
\caption{\baselineskip=7pt Comparison of ATLAS 7 TeV cms energy $W + 1$ jet data for the leading jet $p_T$ distribution and the IR-improved and unimproved exact NLO ME matched parton shower predictions as explained in the text.}
\label{figp2}
\end{figure}
Continuing in this way, we show in Figs.~\ref{figp3} and \ref{figp4} the analogous comparisons~\cite{bsh1} for the ATLAS~\cite{atlas-7-TeV} W+ $\ge$ 2 jet data for the leading and second leading jets, respectively.  
\begin{figure}[h]
\begin{center}
\includegraphics[width=3.5in]{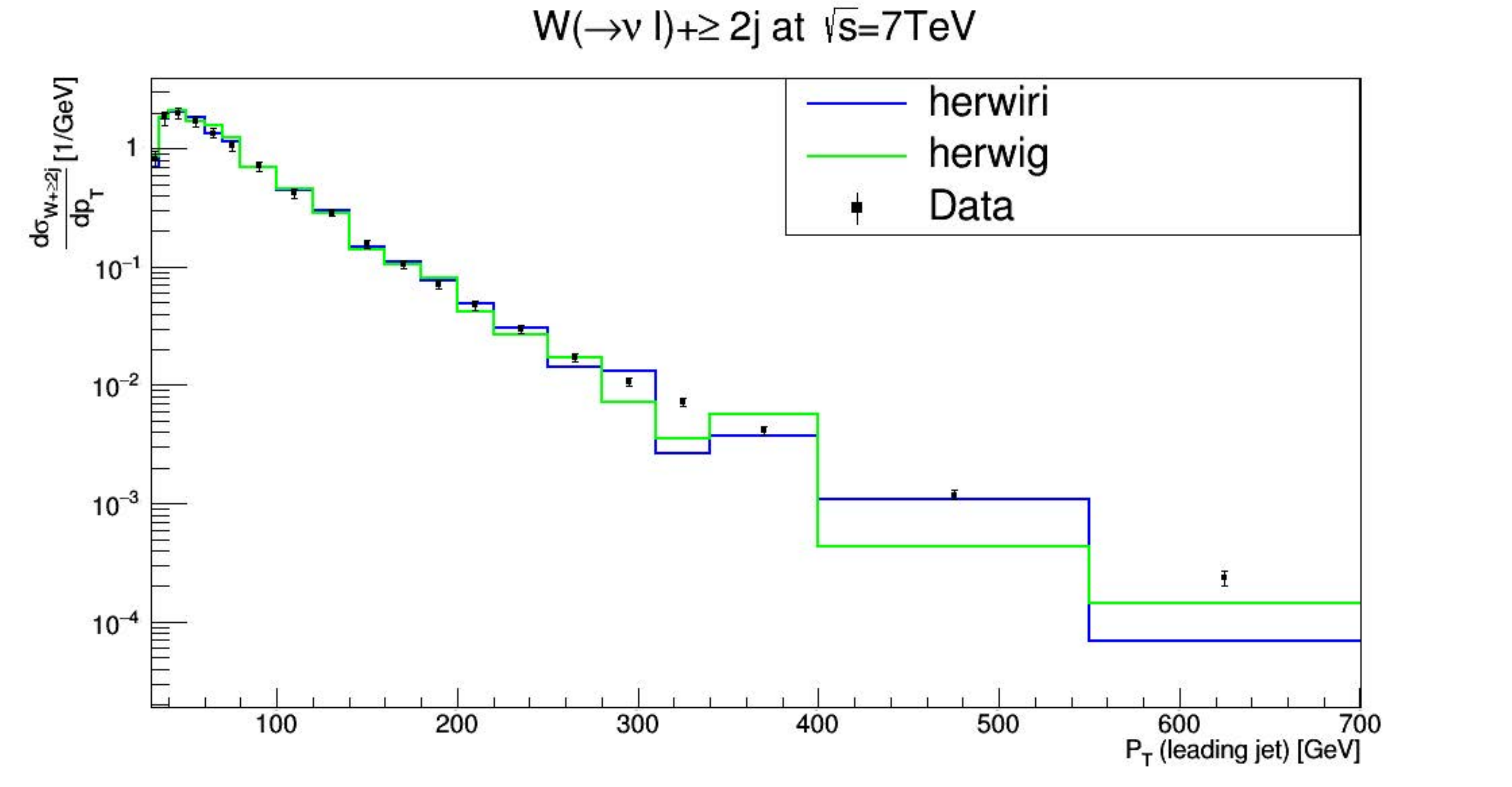}
\end{center}
\caption{\baselineskip=7pt Comparison of ATLAS 7 TeV cms energy $W + \ge 2$ jet data for the leading jet $p_T$ distribution and the IR-improved and unimproved exact NLO ME matched parton shower predictions as explained in the text.}
\label{figp3}
\end{figure}
\begin{figure}[h]
\begin{center}
\includegraphics[width=3.5in]{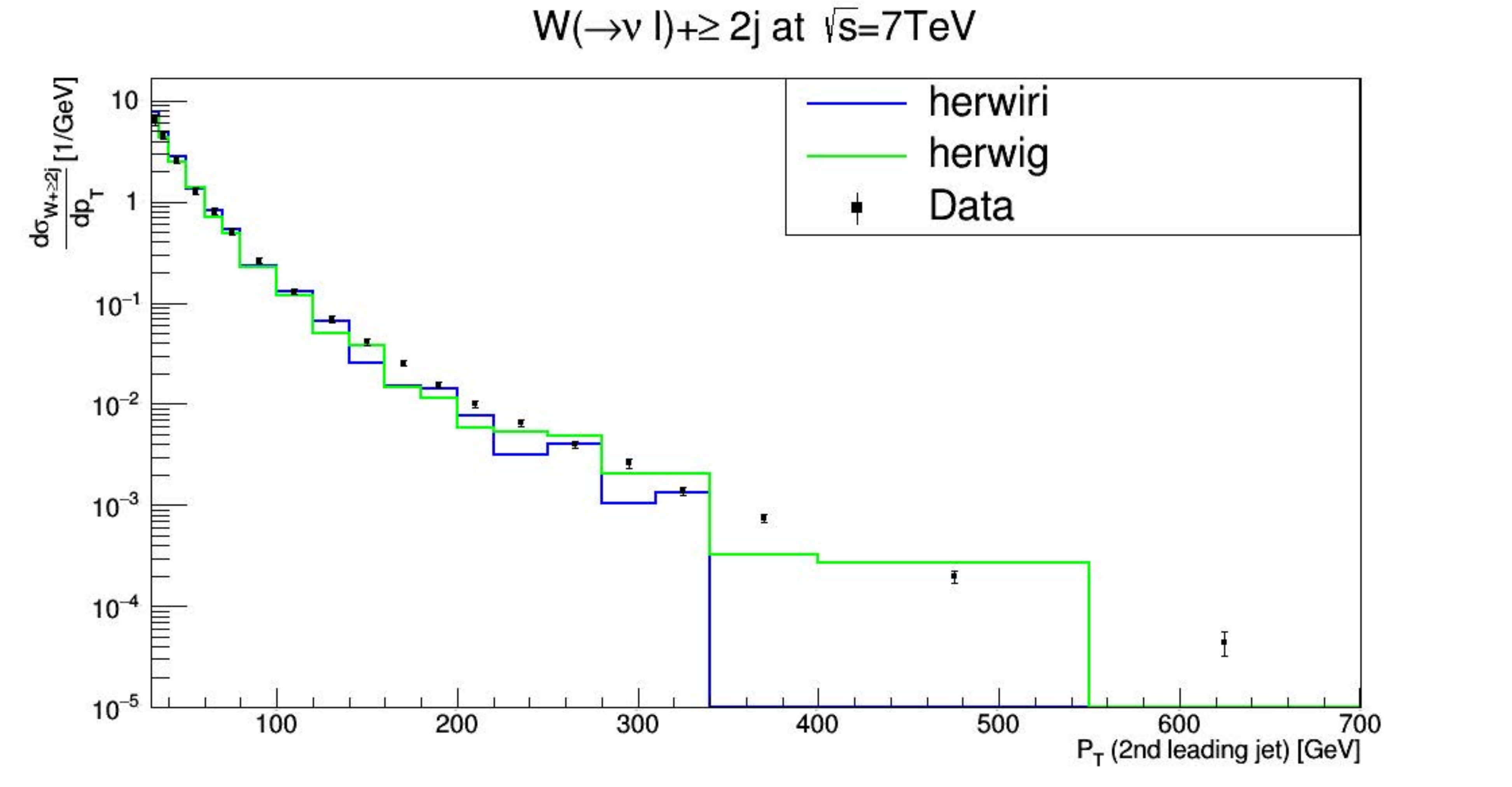}
\end{center}
\caption{\baselineskip=7pt Comparison of ATLAS 7 TeV cms energy $W + \ge 2$ jet data for the second leading jet $p_T$ distribution and the IR-improved and unimproved exact NLO ME matched parton shower predictions as explained in the text.}
\label{figp4}
\end{figure}
In Fig.~\ref{figp5}, we show corresponding comparisons for the leading jet in the ATLAS $W+ \ge 3$ jet data  at 7 TeV. What we see in all of these studies is the in the soft regime below $p_T$ of 180 GeV/c, the IR-improved predictions provide as good or a better description of the data without the need of a 2.2 GeV/c intrinsic Gaussian transverse momentum as it needed in the unimproved (usual) Herwig 6.5.
\begin{figure}[h]
\begin{center}
\includegraphics[width=3.5in]{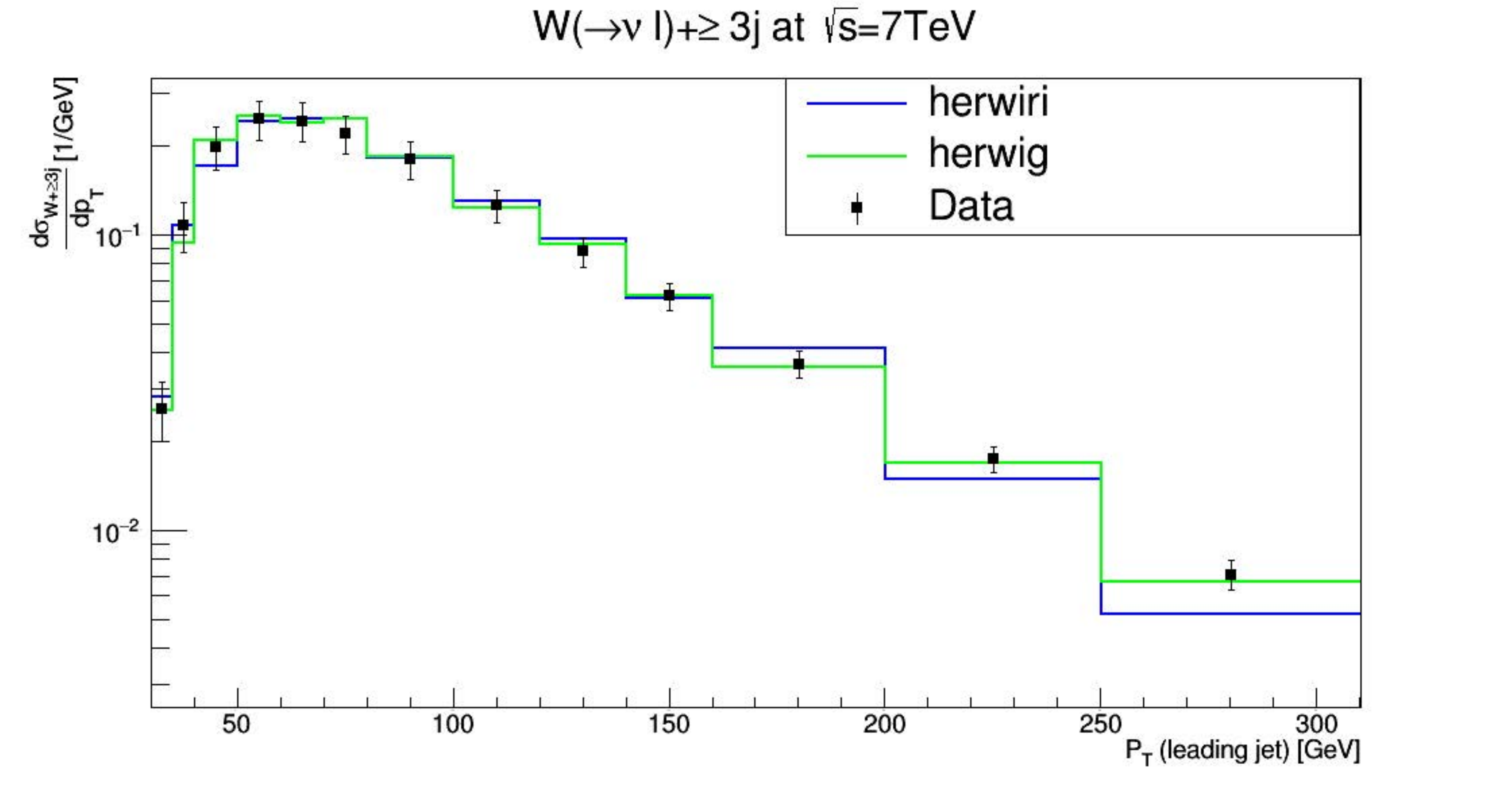}
\end{center}
\caption{\baselineskip=7pt Comparison of ATLAS 7 TeV cms energy $W + \ge 3$ jet data for the leading jet $p_T$ distribution and the IR-improved and unimproved exact NLO ME matched parton shower predictions as explained in the text.}
\label{figp5}
\end{figure}
This extends to the $W+ n$ jets data the corresponding conclusion that was reached by some of us (AM,BFLW and SAY) in studies~\cite{hwri,mcnlo-hwri} of $ Z/\gamma^*$ production at the LHC and FNAL.\par
One question that naturally arises is the effect of IR-improvement on the discovery reach of a standard candle process such as single $Z/\gamma^*$ production at the FCC, to be specific.
Some of us (AM,BFLW and SAY) have investigated the predicted inclusive cross section for $Z/\gamma^*$ as a function of $p_{T,min}$ so that each point in the plot is the cross section
for all events in our simulations with $p_T \ge p_{T,min}$~\footnote{This observable was suggested by M.L. Mangano~\cite{mlm}.}. What we have found is shown in Fig~\ref{figp6}, wherein we plot the respective predictions for the following: MG5\_aMC@NLO/A, A= Herwig6.5, Herwiri1.031, Herwig++ and Pythia8, all with the common renormalization and factorization scale of $M_Z$/2 and all with the common  renormalization and factorization scale of $H_T$/2 (denoted by 'UNFIX' in the legend in the figure) ; MG5\_aMC@NLO/Herwig6.5 and fixed order NLO both with the common renormalization and factorization $M_Z$ for comparison; and, fixed order NLO with the common renormalization and factorization scale $H_T$/2, also for comparison. Here, $H_T$ is the sum of the transverse masses of the final state particles. The cuts are 'ATLAS-like' cuts $\eta_\ell\le 2.4,\; p^\ell_T\ge 20\;{\rm GeV}/c$ and $66\; {\rm GeV}/c^2 \le M_{\ell\bar\ell}\le 116\; {\rm GeV}/c^2$ on the lepton pseudo-rapidity, transverse momentum and invariant mass in an obvious notation. 
\begin{figure}[h]
\begin{center}
\includegraphics[width=100mm]{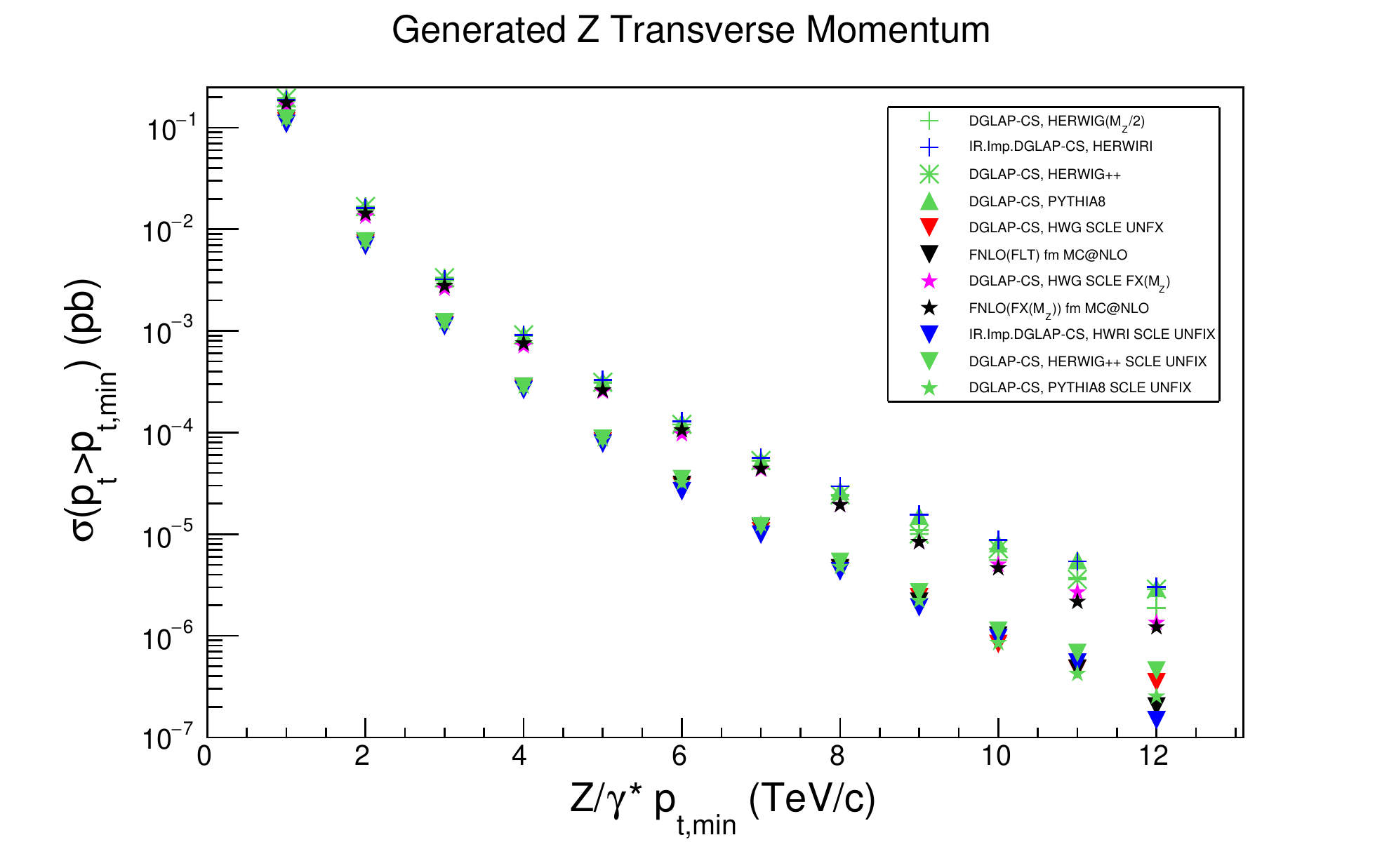}
\end{center}
\caption{\baselineskip=7pt Comparison of various predictions for the FCC discovery plot of the inclusive $Z/\gamma^*$ cross section for evens beyond $p_{t,min}$ as explained in the text and as denoted in the legend.}
\label{figp6}
\end{figure}
We see that the dynamical scale choice makes a big difference in the expectations. The fixed-order NLO results agree with the  MG5\_aMC@NLO/Herwig6.5 results for both of the  scale choices $M_Z$ and $H_T/2$, as it is expected.The dynamical scale choice $H_T/2$  is expected to be more reliable for this observable~\cite{mlm}. The important point is that  IR-improved predictions,
denoted by 'IR.Imp.DGLAP-CS' in the figure, agree with the unimproved ones within the statistical uncertainties.\par
\section{Interplay of IR-Improved DGLAP-CS QCD Theory and Exact ${\cal O}(\alpha^2L)$ CEEX EW Corrections}
The recent ATLAS measurement~\cite{atlas-mw} of $M_W$, which gives the result
$$ 80370\pm 7({\rm stat.})\pm 11({\rm exp. syst.})\pm 14({\rm mod. syst.})\;{\rm MeV}~=~ 80370\pm 19\; {\rm MeV},$$
Z/$\gamma^*$ data used to help establish modeling systematic error denoted here by 'mod. syst.'. Some of us (BFLW, ZAW and SAY) have analyzed the $Z/\gamma^*$ observables used in Ref.~\cite{atlas-mw} from the standpoint of expectations for exact ${\cal O}(\alpha^2L)$ CEEX EW corrections as predicted by {\KK}MC-hh~\cite{kkmchh, kkmchh-atlas,say-thisproc}. The parton shower framework employed in Ref.~\cite{kkmchh-atlas} is that of the unimproved Herwig6.5 MC. Here, we explore the question of the possible role of QCD IR-improvement in the discussion in Ref.~\cite{kkmchh-atlas}.
\par
In Fig.~\ref{figp7}, we show the effect of IR-improvement on the the $Z/\gamma^*$ $p_T$ spectrum for the ATLAS cuts in Ref.~\cite{atlas-mw} using the predictions from 
{\KK}MC-hh/A, A= Herwig6.5, Herwiri1.031. Here, in {\KK}MC-hh/A the 'A' denotes that the respective QCD shower is that from the MC 'A'.
\begin{figure}[h]
\begin{center}
\includegraphics[width=110mm,height=65mm]{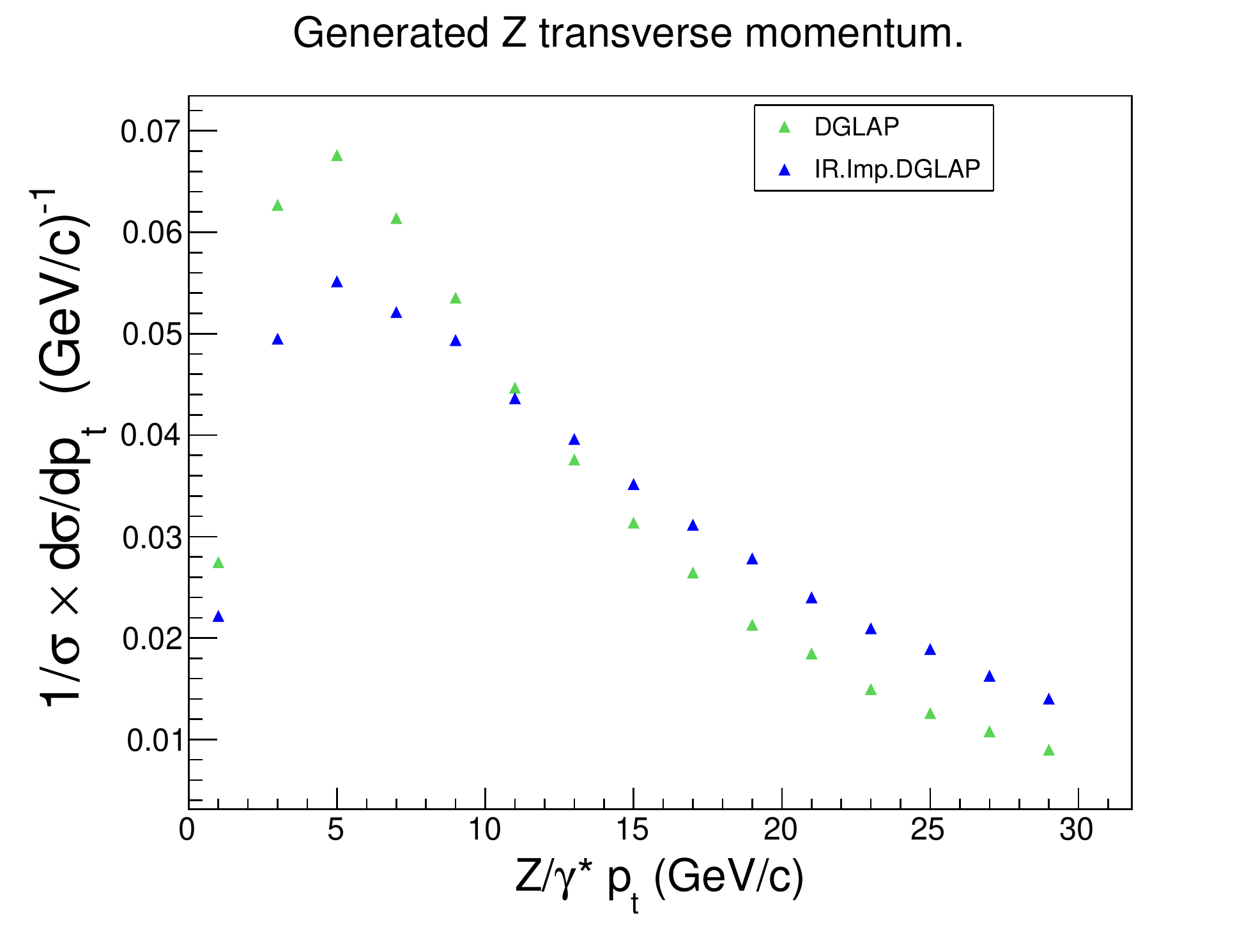}
\end{center}
\caption{\baselineskip=7pt Comparison of {\KK}MC-hh/Herwig(Herwiri) $Z/\gamma^* p_t$ spectra.}
\label{figp7}
\end{figure}
The main feature is the softening  of the spectrum in the soft regime. We are in the process of assessing the corresponding implications for the analysis in Refs.~\cite{atlas-mw,kkmchh-atlas}.\par
\section{Summary}
Precision theory in LHC/FCC physics requires control of both the IR limit, for example, $z\rightarrow 1$ in $q \rightarrow q(z)+\mathfrak{v}(1-z)$, and the collinear limit, for example, $p_T\rightarrow 0$ in $q\bar{q}\rightarrow Z/\gamma^*(\vec{p}_T)+\mathfrak{v}(-\vec{p}_T)$, where in the context of $QED\otimes QCD$ resummation we have $\mathfrak{v}=\gamma,\; G.$ In the latter process  example, both $q$ and $\bar{q}$ have $p_T=0$. We now have control over both limits in QCD in MG5\_aMC@NLO/Herwiri1.031, for example, and in $QED\otimes QCD$ in {\KK}MC-hh/Herwiri1.031, as another example.
We continue to point out that some New Physics at both the LHC and the FCC may obtain therefrom.\par
\section*{Acknowledgments} One of us (BFLW) would like to thank Prof. G. Giudice for the support of the CERN-TH Department while a part of this work was done. We thank Prof. S. Jadach and Dr. S. Majhi for helpful discussions and inputs.

\end{document}